# Optical detection of NMR J-spectra at zero magnetic field

M. P. Ledbetter[1], C. W. Crawford[2], A. Pines[2,3], D. E. Wemmer[2], S. Knappe[4], J. Kitching[4], and D. Budker[1,5]

[1]*Department of Physics, University of California at Berkeley, Berkeley, California 94720-7300 USA*, [2]*Department of Chemistry, University of California, Berkeley, California 94720 USA*, [3]*Materials Sciences Division, Lawrence Berkeley National Laboratory, Berkeley, California 94720 USA*, [4]*Time and Frequency Division, National Institute of Standards and Technology, 325 Broadway, Boulder, Colorado, 80305 USA*, [5]*Nuclear Science Division, Lawrence Berkeley National Laboratory, Berkeley, California 94720,USA*

**Scalar couplings of the form $J\mathbf{I}_1 \cdot \mathbf{I}_2$ between nuclei impart valuable information about molecular structure to nuclear magnetic-resonance spectra. Here we demonstrate direct detection of J-spectra due to both heteronuclear and homonuclear J-coupling in a zero-field environment where the Zeeman interaction is completely absent. We show that characteristic functional groups exhibit distinct spectra with straightforward interpretation for chemical identification. Detection is performed with a microfabricated optical atomic magnetometer, providing high sensitivity to samples of microliter volumes. We obtain 0.1 Hz linewidths and measure scalar-coupling parameters with 4-mHz statistical uncertainty. We anticipate that the technique described here will provide a new modality for high-precision "J spectroscopy" using small samples on microchip devices for multiplexed screening, assaying, and sample identification in chemistry and biomedicine.**

Nuclear magnetic resonance (NMR) endures as one of the most powerful analytical tools for detecting chemical species and elucidating molecular structure. The fingerprints for identification and structure analysis are chemical shifts and scalar couplings (*1,2*) of the form $J\mathbf{I}_1 \cdot \mathbf{I}_2$. The latter yield useful information about molecular spin topology, bond and torsion angles, bond strength, and hybridization. NMR experiments are conventionally performed in high magnetic fields, requiring large, immobile, and expensive superconducting magnets. However, detection of NMR at low magnetic fields has recently attracted considerable attention in a variety of contexts, largely because it eliminates the need for superconducting magnets. One-dimensional (*3*) and two-dimensional (*4*) spectroscopy have been demonstrated in the Earth's magnetic field using inductive detection, J-resolved spectra have been detected with superconducting quantum interference device (SQUID) magnetometers in ~μT fields (*5*), and atomic magnetometers have been used to perform one-dimensional spectroscopy (*6,7,8*) and for remote detection of magnetic resonance imaging (*9,10*) in low magnetic fields. Field cycling has been used in the past to observe heteronuclear scalar coupling in a zero field environment (*11,12*), however the practice has not become widely used as it entails cumbersome shuttling of a sample in and out of a high-field magnet.

Here we demonstrate direct detection of hetero- and homonuclear scalar coupling in magnetic zero-field using an optical atomic magnetometer. We show that characteristic functional groups have distinct spectra, with straightforward interpretation for molecular structure identification, allowing extension to larger molecules and to higher dimensional Fourier NMR spectroscopy. A magnetically shielded, zero-field environment provides high absolute field homogeneity and temporal stability, allowing us to obtain 0.1-Hz linewidths without using spin echoes, and to determine scalar coupling parameters with a statistical uncertainty of 4 mHz.

The use of atomic magnetometers yields greatly improved sensitivity compared to inductive detection at low or zero fields because they sense magnetic field directly, rather than the time derivative of flux through a pickup coil. Furthermore, in contrast to SQUIDs, atomic magnetometers do not require cryogenics. We achieve efficient coupling to small samples by making use of millimeter-scale magnetometers (*13*) manufactured using microfabrication techniques (*14*). These factors allow us to work with an 80-μL detection volume, 25 and 6000 times smaller than the quantities used in the Earth-field studies of Refs. (*3*) and (*4*), respectively. We also use magnetic shielding, which permits operation in a laboratory environment, where perturbations to the Earth's magnetic field may limit the magnetic field homogeneity and stability.

Operation at zero field eliminates the chemical shift but retains substantial analytical information in simplified spectra determined by both heteronuclear and homonuclear scalar couplings. The $^{13}CH_3$ group provides an example of the simplification afforded by spectroscopy in a zero-field environment: the Earth's field spectrum consists of eight lines (*15*), while, as we show here, the zero-field spectrum consists of just two lines, without loss of spectral and analytical information. This will facilitate controllable extension into multidimensional spectroscopy (*16*) with the incorporation of zero-field decoupling and recoupling sequences (*17,18*).

At zero magnetic field, the Hamiltonian for a network of spins coupled through scalar interactions is

$$H_J = \hbar \sum J_{jk} \mathbf{I}_j \cdot \mathbf{I}_k , \qquad (1)$$

where the sum extends over all spin pairs and $J_{jk}$ is the J-coupling parameter for spins *j* and *k*. The observable in our experiment is the *z* component of the magnetization of the sample (see supporting online material),

$$M_z(t) = \hbar n \mathrm{Tr}\left(\rho(t)\sum_j \gamma_j I_{j,z}\right), \tag{2}$$

where $n$ is the number density of molecules, $\gamma_j$ is the magnetogyric ratio of the $j$-th spin, and $\rho(t)$ is the density matrix. The temporal evolution of an arbitrary system of spins can be determined by diagonalizing the Hamiltonian to find the eigenstates $|\varphi_a\rangle$ and eigenvalues $E_a$, and expressing the initial density matrix as a sum of the operators $|\varphi_a\rangle\langle\varphi_b|$, each of which evolves as $e^{i\omega_{ab}t}$, where $\omega_{ab} = (E_a - E_b)/\hbar$.

Because $I_{j,z}$ are vector operators with magnetic quantum number zero, observable coherences are those between states that differ by one quantum of total angular momentum **F**, $|\Delta F|=1$ with $\Delta M_F = 0$. This selection rule can be used for prediction of the positions of peaks and for interpretation of spectra. For instance, consider the case of $^{13}$CH$_N$, where the J-coupling $J_{CH}$ between all N heteronuclear pairs is identical. Since the protons are all equivalent, the homonuclear J-couplings can be ignored (*2*). Denoting the total proton spin by **K** and the $^{13}$C spin by **S**, Eq. (1) can be rewritten $H_J = \hbar J_{HC} \mathbf{K} \cdot \mathbf{S}$, which has eigenstates $|F^2, K^2, S^2, F_z\rangle$ with eigenvalues

$$E_{F,K} = \hbar \frac{J_{HC}}{2}\left[F(F+1) - K(K+1) - S(S+1)\right]. \tag{3}$$

The selection rules above yield the observable quantum-beat frequencies $\omega_K = (E_{K+1/2,K} - E_{K-1/2,K})/\hbar = J_{HC}(K+1/2)$ for $K \geq 1/2$. For the methyl group, $^{13}$CH$_3$, we expect two lines, one at $J_{HC}$ and another at $2J_{HC}$, corresponding to coupling of the $^{13}$C nucleus with the proton doublet or quadruplet states. For the methylene group, $^{13}$CH$_2$, a single line at $3J_{HC}/2$ is expected due to coupling with the proton triplet state. In more complicated molecules, homonuclear couplings or higher-order effects of heteronuclear couplings can result in a splitting of the lines – however, the positions of the multiplets can be determined by the above argument.

A schematic of our zero-field spectrometer is shown in Fig. 1. A syringe pump cycles fluid between a prepolarizing volume and the 80-μL detection volume adjacent to an optical-atomic magnetometer. The prepolarizing volume is placed in a compact (5 cm × 5 cm × 10 cm) 1.8-T Halbach array. A set of magnetic shields and coils surrounding the magnetometer and detection volume create a zero-field environment to a level of 0.1 nT. The central component of the magnetometer is a vapor cell, with inner dimensions 2.7 mm × 1.8 mm × 1 mm, containing $^{87}$Rb and 1200 Torr of $N_2$ buffer gas, fabricated using the techniques described in Ref. (*14*). The atomic magnetometer operates in the spin-exchange relaxation-free (SERF) regime (*19*), in which relaxation of the alkali polarization due to spin-exchange collisions is eliminated. As in Ref. (*13*), we use a single circularly polarized laser beam tuned to the center of the pressure broadened Rb D1 transition, propagating in the *x* direction, to optically pump and probe the alkali polarization. The configuration of the magnetometer (see supporting online material) is such that it is primarily sensitive to the *z* component of the magnetic field, or sample magnetization. Inset (a) in Fig. 1 shows the response of the magnetometer to a small oscillating test field as a function of frequency. Inset (b) in Fig. 1 shows the sensitivity of the magnetometer (the sharp peaks are for calibration) after normalizing the measured noise and calibration signals by the frequency response of the magnetometer, yielding a noise floor of about 200 fT/√Hz, flat from about 3 Hz to 300 Hz. For more details on the operation of the magnetometer, see the supporting online material.

Data presented in this work is acquired as follows: Polarized fluid flows into the detection region, and at *t*=0, flow is halted and a pulse of DC magnetic field is applied in the *y* direction with magnitude $B_1$ and duration $T_p$. This rotates the proton and $^{13}$C spins by different angles due to the different magnetogyric ratios, placing the spin system into a superposition of eigenstates of the J-coupling Hamiltonian, Eq. (1). The ensuing quantum beats lead to a time-dependent magnetization, the *z* component of which is detected by the atomic magnetometer. The transfer of the sample from high field to zero field is adiabatic as no quantum beats are observed without application of an excitation pulse. Adiabatic transfer results in an equilibration of the spin-temperature parameters of the two species via the J-coupling interaction, while preserving the $z$ component of total angular momentum (20). This is the initial condition for simulations presented below.

Measurements on methanol, $^{13}CH_3OH$ are presented in Fig. 2 for a pulse area $\alpha = B_1 T_p (\gamma_H - \gamma_C) = 2.4$ rad ($T_p = 0.66$ ms). The signal in the time domain after averaging 11 transients is shown in Fig. 2(a). There is a large, slowly decaying component of the signal due to the relaxation of static components of the total magnetization, as well as a smaller, high frequency component due to scalar coupling. Overlaying the data is a decaying exponential (red trace) with time constant $T_1 = 2.2$ s. In displaying these data, we first subtracted the decaying exponential, filtered the remaining signal with a pass band between 120 and 300 Hz and then added the decaying exponential to the filtered data. This eliminates transients at the beginning and end of the data set due to the digital filter. The Fourier transform of the signal is shown in Fig. 2(b) after correcting for the finite bandwidth of the magnetometer, revealing a simple structure consisting of two peaks (offsets inserted for visual clarity). This spectrum is in agreement with the discussion of $^{13}CH_3$ given above, assuming that the homo- and heteronuclear coupling of the OH group are averaged to zero under rapid chemical exchange. Independently fitting the low- and high-frequency portions of the data to complex Lorentzians yields central frequencies $\nu_1 = 140.60$ Hz and $\nu_2 = 281.09$ Hz with linewidths (half-width at half-maximum) $\Delta\nu_1 = 0.10$ Hz and $\Delta\nu_2 = 0.17$ Hz.

The dependences of the amplitudes of the low- and high-frequency peaks on the pulse area are shown by triangles and squares, respectively in Fig. 3 (a). The black and red lines overlying the data are theoretical predictions, in agreement with the data. Note that the dependences are not quite sinusoidal because the measured amplitude depends on the relative orientation of the two nuclei as well as the spatial orientation of the total angular momentum, both of which are affected by the pulse.

In order to determine the stability of the J-coupling measurement, we acquired a series of 100 transients following the application of a pulse with area $\alpha = 2.4$ rad, the first maximum of the response in Fig 3(a). The raw data were binned into sets of ten, averaged, Fourier transformed, and fit to complex Lorentzians. The position of the low- (triangles) and high- (squares) frequency peaks are shown as a function of bin number in Fig. 3(b). The mean frequencies of each peak are indicated by the solid lines overlying the data with $v_1 = 140.566(4)$ Hz and $v_2 = 281.082(3)$ Hz. These values are in agreement with the value found in the literature of $J_{HC}=140.6$ Hz for methanol (*15,21*) (presented without explicitly stated uncertainty). However, these data deviate slightly from the $^{13}CH_3$ model discussed above because $v_2/2$ differs from $v_1$ by about 25 mHz. We suspect that this small shift is the result of residual coupling to the OH group, and simulation indicates that it would require a coupling of only 0.4 Hz to produce a shift of this magnitude and sign. The statistical uncertainties in our measurements are orders of magnitude smaller than the range of frequencies associated with J-couplings, providing a sensitive probe for subtle differences in chemical structure.

As mentioned above, homonuclear J-coupling between equivalent spins cannot be observed. In high-field NMR experiments, this is often overcome by differences in chemical shift between different functional groups. At low to zero magnetic fields where chemical shifts are unresolved or non-existent, homonuclear non-equivalence can occur through different heteronuclear J-coupling environments. (*15*) For example, in ethanol 1, $^{12}CH_3$-$^{13}CH_2$-OH, or ethanol 2, $^{13}CH_3$-$^{12}CH_2$-OH, the protons in the methyl and methylene groups couple to the $^{13}C$ nucleus differently, yielding observable effects due to homonuclear J-coupling. Figure 4 shows experimental spectra for ethanol 1 and ethanol 2, obtained after averaging 210 and 475 transients, respectively. Simulated spectra, presented below the data, are in agreement with experiment. In the simulations, we use the values of coupling constants obtained from high-field measurements, which, for ethanol 1 are $J_{HC}^{(1)} = 140.4\,\text{Hz}$, $J_{HC}^{(2)} = -4.6\,\text{Hz}$ and $J_{HH}^{(3)} = 7.1\,\text{Hz}$ and for ethanol 2 are $J_{HC}^{(1)} = 125.2\,\text{Hz}$, $J_{HC}^{(2)} = -2.4\,\text{Hz}$ and $J_{HH}^{(3)} = 7.1\,\text{Hz}$ (21,22), where the superscript denotes the number of bonds separating the interacting nuclei. These spectra can be interpreted as follows: The Hamiltonian is dominated by the one-bond heteronuclear J-coupling. Hence, neglecting any other couplings, for ethanol 1, one expects a single peak at $3J_{HC}^{(1)}/2$ due to coupling between the $^{13}$C nucleus and the triplet proton state of the methylene group. In ethanol 2, one expects two peaks at $J_{HC}^1$ and $2J_{HC}^1$ due to coupling between the $^{13}$C nucleus and the doublet or quadruplet states of the protons on the methyl group. Homonuclear couplings and two-bond heteronuclear couplings result in a splitting of these peaks, as well as the appearance of a set of peaks at low frequencies.

In the present work, the magnetometric sensitivity is about 200 fT/√Hz, with a vapor cell volume of about 4.8 mm$^3$. Laser intensity fluctuations are the dominant source of noise and are about a factor of 50 larger than photon shot noise. A straightforward path to improved sensitivity would be to incorporate a second, low noise laser, and monitor optical rotation, which would cancel common mode noise. Fundamentally limiting the sensitivity of an atomic magnetometer is spin-projection noise (23), and in Ref. (8) we estimate that, for millimeter-scale vapor cells with optimal values of parameters such as light power, cell temperature, and buffer gas pressure, spin-projection noise is on the order of 0.1 fT/√Hz, indicating that there is still a great deal of room for improved magnetometric sensitivity. Hyperpolarization techniques such as dynamic nuclear polarization (24) or parahydrogen-induced polarization (25) can also be employed to yield much larger signals, making possible the detection of natural-abundance samples.

In conclusion, we have demonstrated direct detection of pure J-coupling NMR at zero magnetic field using an optical atomic magnetometer. For characteristic functional groups, such as $^{13}CH_3$, the zero-field spectrum is simpler than Earth-field spectra (*15*) while retaining all information about the J-coupling network. We obtain linewidths as low as 0.1 Hz, measure heteronuclear J-coupling constants with 4-mHz statistical uncertainty and clearly observe homonuclear J-coupling. Zero-field relaxation rates can also easily be measured in our experiment with only a single pulse. The sensitivity is sufficient to obtain simple spectra from 80 μL of fluid in a single shot. Further optimization of magnetometric sensitivity and geometry will yield improved performance with detection volumes at the level of 1 μL. We anticipate that the technique described here will find wide use in analytical chemistry. One particular application we envision for the present technique is in monitoring changes of scalar couplings in enzyme catalyzed reactions. Applications to multiplexed screening, assaying and identification of samples from chemistry to biomedicine (*26*) with mobile, miniaturized devices are also envisaged.

**Figure Captions:**

**Figure 1:** Experimental setup. A syringe pump pushes fluid from a reservoir inside a 1.8-T Halbach array, through the 80-μL detection volume adjacent to an $^{87}Rb$ alkali-vapor cell. The vapor cell and detection volume are housed inside a set of magnetic shields. Circularly polarized light from an external-cavity diode laser at the D1 resonance is used to optically pump and probe the alkali spin polarization. A set of coils inside the magnetic shields is used to zero the residual magnetic field and apply pulses to the sample. An oven heats the cell to 170°C to maintain sufficient alkali vapor density. Insets (a) and (b) show the response of the magnetometer to test fields of varying frequency and the noise floor of the magnetometer, respectively.

**Figure 2**: Raw signal (a) and Fourier transform (b) obtained following an excitation pulse with area $B_1 T_p (\gamma_H - \gamma_C) = 2.4$ rad. In the top panel, the smooth red curve overlaying the data is a decaying exponential with a time constant $T_1 = 2.2$ s. The real and imaginary parts of the spectrum are represented in (b) by the black and red traces, respectively. The low- (high-) frequency peaks correspond to the coupling of the $^{13}$C nucleus with the doublet (quadruplet) states of proton angular momentum.

**Figure 3:** (a) Triangles and squares show the dependence of the amplitude of the low- and high-frequency resonances in $^{13}$C enriched methanol on pulse area, respectively. The solid lines overlaying the data are theoretical predictions. (b): The center of the low (triangles) and high (squares) frequency resonances as a function of bin number, each bin consisting of the average of 10 transients. From these data, we determine the mean value of the central frequency for the two peaks to be 140.566(4) and 281.082(3) Hz, as indicated by the solid lines overlaying the data.

**Figure 4:** Experimental and simulated zero-field NMR spectra for ethanol 2 (top panel), $^{13}CH_3$-$^{12}CH_2$-OH and ethanol 1, $^{12}CH_3$-$^{13}CH_2$-OH. To the extent that signal is above the noise level, experiment and simulation are in agreement. The positions of the multiplets are determined by the one-bond heteronuclear J-coupling and the splittings within the multiplets are due to homonuclear J-coupling and two-bond heteronuclear J-coupling.

**Acknowledgements**

The authors sincerely appreciate useful discussions with E. L. Hahn, S. M. Rochester, L.-S. Bouchard, and V. M. Acosta. This work was supported by the ONR-MURI grant No. FD-N00014-05-1-0406, by the Director, Office of Science, Office of Basic Energy Sciences, Nuclear Science Divisions, of the U.S. Department of Energy under contract DE-AC03-76SF00098, and by the Microsystems Technology

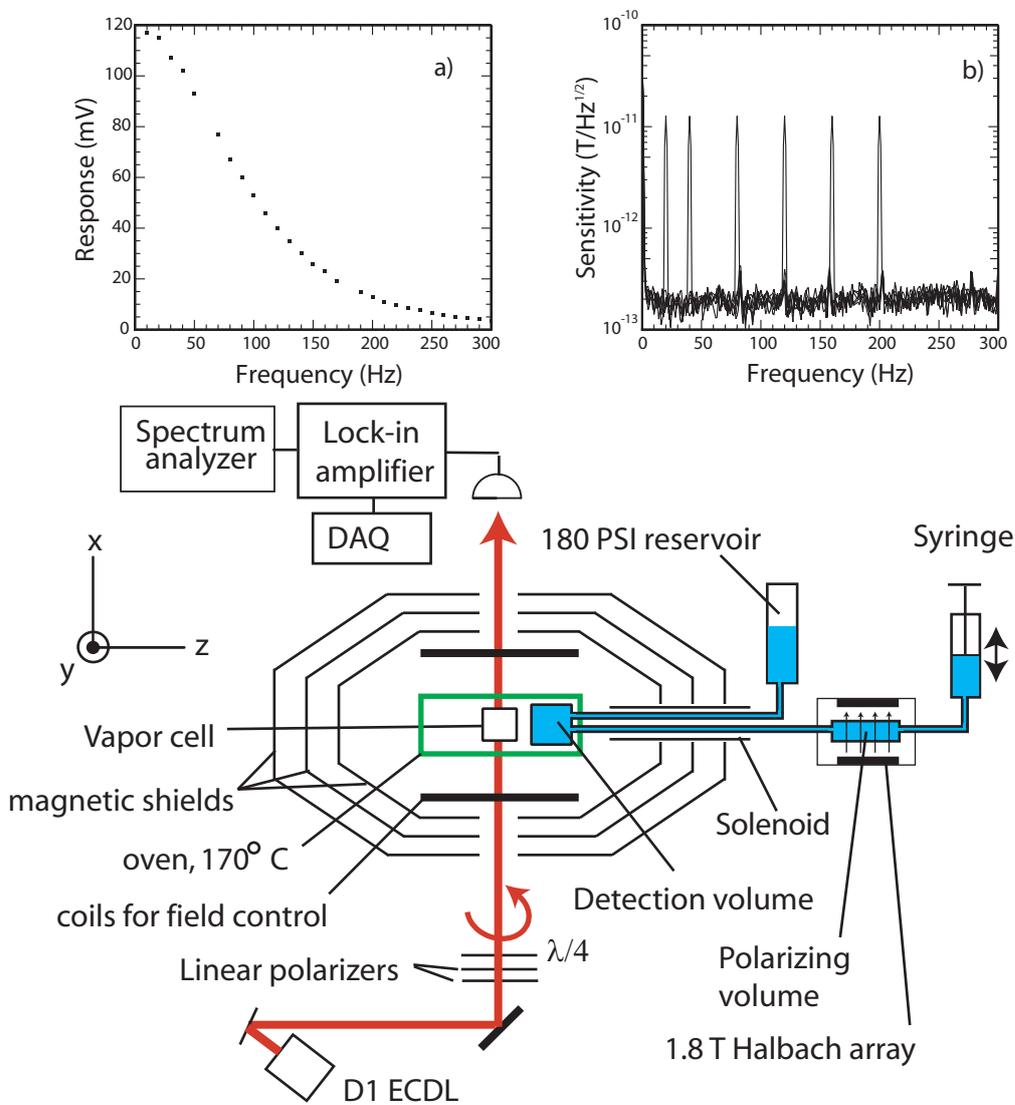

**Figure 1**

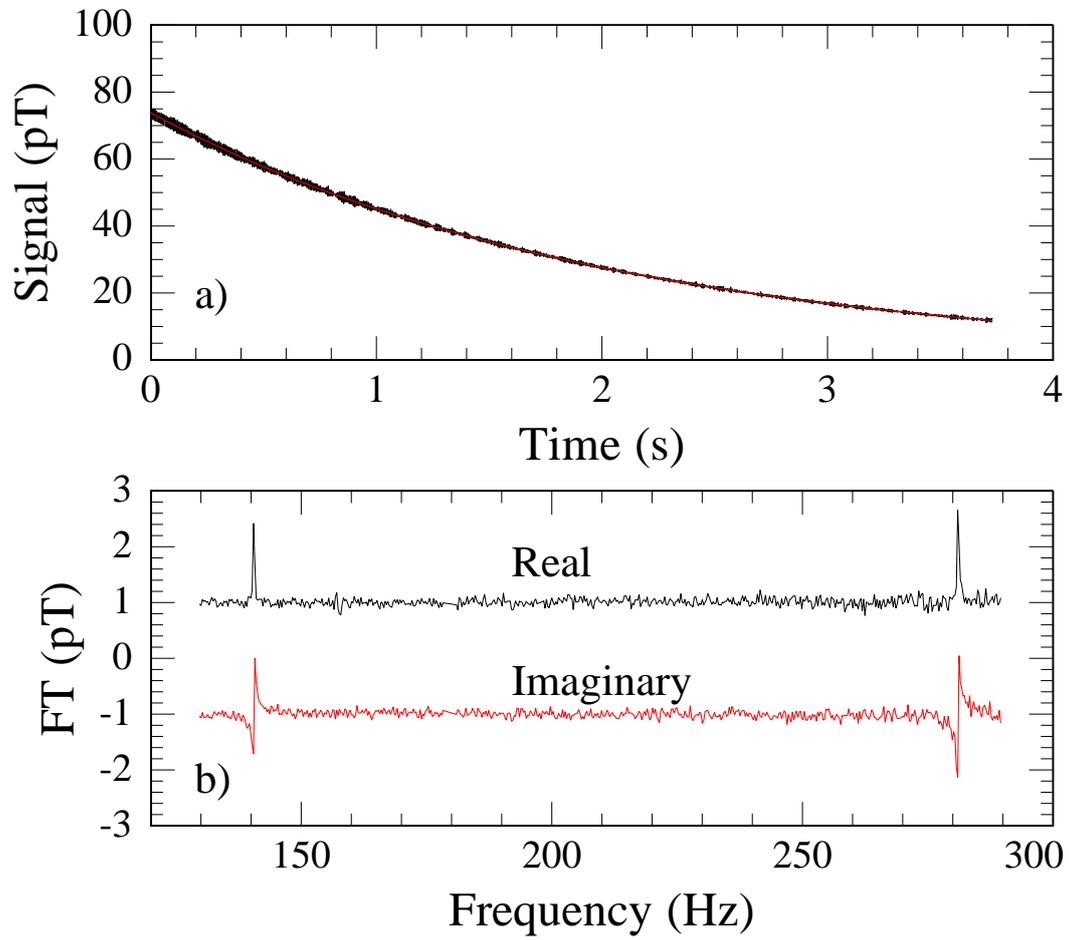

**Figure 2**

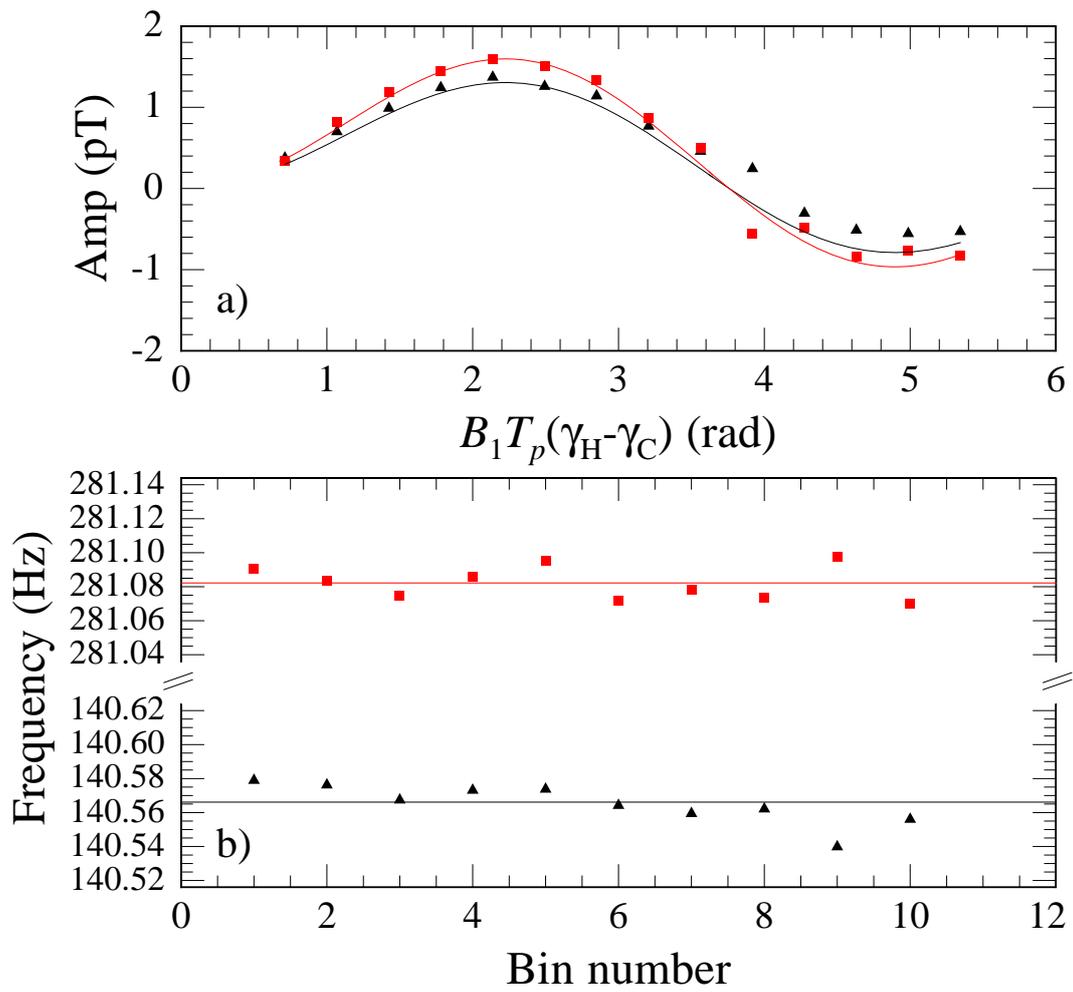

**Figure 3**

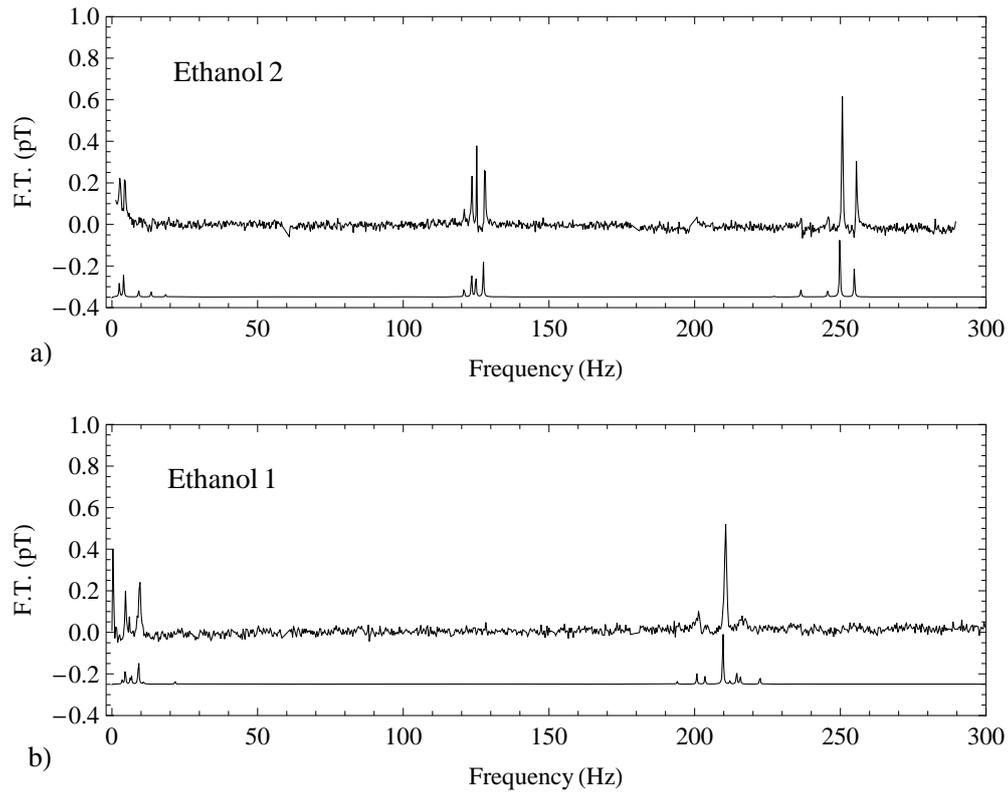

**Figure 4**